# Towards a theory of consciousness: Proposal for the resolution of the homunculus fallacy with predictions[*]

A. Lőrincz and G. Szirtes

Department of Information Systems, Eötvös University of Sciences, Budapest, Hungary

Abstract

In this paper we argue that no forms of Turing test are either necessary or sufficient to establish if a machine is conscious or not. Furthermore, from a modeling point of view, the problem is that the Turing test does not really provide testable predictions. We believe that the model structure should explain the function (of consciousness). We argue that the cornerstone of any model on consciousness is to (partly) overcome the obstacle of the homunculus fallacy about the use of representations. In this contribution a possible solution is suggested, which makes use of reflexive architectures. The emerging computational constraints on the proposed architecture have lead to testable predictions on the dynamical behavior of the biological substrate. Interestingly, these predictions are in good agreement with recent experimental observations.

## Introduction

Turing's famous proposal (Turing, 1950) on a general criterion set for modeling cognition has begun a new chapter in discovering the nature of consciousness, the problem, which is as old as the philosophy itself. Turing's work can be interpreted as an intention to bridge the gap between the philosophy of the mind and the so-called hardcore sciences (e.g., computational neuroscience, neurobiology, etc.). However, instead of facilitating and directing experimental research, the presumed generality of the Turing test gave rise to new and disturbing questions, many of which, from a modeling point of view, lead us into deadlocks.

We admit that attacking the Turing test (e.g. Searle, 1980) and defending it (see e.g. Harnad 2000a, 2000b, 2003) are both entertaining and useful in clarifying the logic behind the philosophy of intelligence and/or consciousness and/or thinking. We are afraid, however, that these routes cannot lead to testable model construction. This is why we feel inspired to chip on the conversation. We hope that our previous works on neurobiological modeling qualifies us for such contribution. Our contribution does not pretend to provide a direct solution to the problem of consciousness. Instead, our remarks are meant to be a call to extend the objectives of cognitive research.

## Problems of Turing tests

In this paper we principally focus on Stevan Harnad's thoughts from a pragmatic standpoint, because he addresses many exciting issues and aims to highlight their convergence. From our modeling perspective, he argues for the interrelatedness of the mind – body problem, the so-called 'other minds' problem and the problem of symbol grounding. Summarizing Harnad's argument (Harnad 2003, 2000b), we have found the following statements important for our reasoning: i, While the presence of consciousness in others (from human beings to artificial creatures to enter the club of conscious things) is an ontological question, having the knowledge about it is an epistemic problem that is (following Descartes) only our own belief (based on, e.g., the experience of others' *behavior*) can help us in deciding on others' club membership (i.e., epistemic inference on ontological problem). ii, The Turing test still remains the best device to support such an experience based decision.

The weakness of the original form of the Turing test (T2) has been demonstrated by Searle's (1980) Chinese Room argument. As a partial solution, other levels of Turing tests have been introduced. For example, T4 corresponds to a system, which has indistinguishable internal functioning (even at the neuro-molecular level), while T5 is indistinguishable in "every empirically discernible respect" (Harnad, 2000b). The intermediate level of T3 scales up T2 to full performance capacity to pass the Total Turing Test (T3) (see, e.g., in Harnad, 2000a, b).

In our view, both T3 and T4 (T5 is omitted for the sake of simplicity) have the same inescapable drawback: they are inherently anthropomorphic, only the mimicking capabilities are different. In other words, passing Turing test is not a *sufficient condition* for recognizing or defining consciousness.

We claim that Turing tests are not *necessary conditions* either. An illuminating but somewhat controversial example is the case of patients awoken from coma, who can report conscious-like sensations and mental processes during their coma state, which are not accessible for any Turing test. Young children before acquiring the ability of language would be another example, because they would also fail in a communication based Turing test. Thirdly, although there are famous demonstrations that some monkeys can use some hand signs based on American Sign Language and could pass certain items of T2-T4 Turing tests, neither of these tests can tell us, if lower level primates are conscious or not. To escape from this seemingly vicious circle of a neither necessary nor sufficient test, we suggest looking for other, possibly less anthropomorphic constraints on functional modeling with explanation power.

As a first step, by rephrasing the problem with Harnad's wording, we think that the issue at hand is how a "ghost in a machine" could convince itself about the presence or absence of "a ghost in another machine".

Our proposal is to find key modeling issues, which *might* lead to a *functional* explanation of the "ghost". When functional *models* are investigated, one is forced to deal with internal functioning, which is inherently tied up with the use of *representations*. However, the concept of using representations is not without problems. The main attack against such representation has been clearly described by, e.g., Dennett (1991) and Searle (1992) in the form of the so-called "homunculus" fallacy. According to the fallacy, the internal representation in any information processing system is meaningless without an interpreter. The paradox claims that all levels of abstraction require at least one further level containing the corresponding interpreter. The interpretation – according to the fallacy – is just a new transformation and we are trapped in an endless recursion. An intriguing property of the fallacy is its generality: it is hard to think of a conscious being, which does not have any form of any interpreting function. Thus, the fallacy is not restricted to humans.

It is our firm belief that *any model* targeting conscious mental processes, such as declarative memory, decision-making and planning (and feelings if you like), could be questioned by the arguments of the fallacy. That is, a constructive route to find the "ghost in the machine" is to resolve the fallacy *first*. Our thinking is best expressed by the words of Albert Szent-Györgyi (1951), the famous Hungarian Nobel Laureate: "There is no real difference between structure and function; they are the two sides of the same coin. If structure does not tell us anything about function, it means we have not looked at it correctly."

## A resolution of the fallacy

First, we note that there can be more than one route to resolve the fallacy. For example, along the line of the classical black box modeling, the fallacy does not arise at all (see, e.g., Dennett (1991)). The price to pay is that black box modeling cannot provide structural explanation and must resort in the Turing tests.

We claim that the paradox stems from vaguely described procedure of 'making sense'. The fallacy arises by saying that the internal representation should make sense. To the best of our knowledge, this formulation of the fallacy has not yet been questioned except in our previous works (Lőrincz, 1997, Lőrincz et al. 2002a, b). The fallacy was turned upside down by changing the roles: Not the internal representation but the *sensory input,* e.g., retinal pattern, or its transformed forms, should make sense: The *input makes sense* if the same (or similar) inputs have been experienced *beforehand* and if the input can be derived or regenerated by means of the internal representation (Lőrincz, 1997, Lőrincz 1998, Lőrincz et al., 2002a, b, c). According to this approach the internal representation interprets the input by (re-) constructing it.

The idea behind this approach is to execute the infinite recursion in a *finite* architecture. The change of the roles gives rise to a reconstructing loop structure. The loop has two constituents; the top and the bottom. The top part contains the internal representation that, in turn, generates the reconstructed input via the top-down transformation. The bottom part computes the difference between the actual input and the reconstructed input. This difference, the reconstruction error is then used to correct the internal representation via the bottom-up transformation, which generates (modifies) the reconstructed input and so on. This is a finite architecture with a converging, but – in principle – endless iteration and the fallacy is simplified to the problem of stability and convergence. It may be important to note that this route has nothing to do with mirroring (the external world). The input to the mirror and the mirror image differ in their material qualities and the mirror has no tool to compare the two and to engage in any iteration to make corrections. One might say that the internal representation, which reproduces the input, is a (spatio-temporal) *model* in a general sense: it *predicts* and reproduces (internal and external) sensory information (Lőrincz et al., 2002a, b, c).

There are relatively strong (mathematical and computational) constraints on how such a reconstruction network should work. These constraints severely restrict our freedom in building such architectures (Lőrincz 2002b).

A few corollaries of the resolution

We followed the aforementioned constructive route and derived a model (Lőrincz and Buzsáki, 2000, Lőrincz et al., 2000a). The model has some emerging mathematical properties For example, successful reconstruction trivially requires at least two things:
1. The information content of the input should be accessible for the internal representation
2. The noise content of the input should not be reconstructed and, in turn, it should not be available for the internal representation

The first request can be ensured by the maximization of information transfer of the bottom-up transformation – as it has long been suggested by Attneave (1954) and Barlow (1961). To achieve this, the bottom-up processing channels should be adjusted on the base of the arriving inputs. Loosely speaking, Point 2 says that (a) information and noise should be distinguished and (b) the bottom-up filtering should cancel the noise content. Distinction between information and noise is based on the degree of their compressibility. The concept of compressibility is related to the recognition of any organized structure in the input. If the given architecture is able to *compile* and interpret the input then it can also provide a more compact description. While a representation can be even more complex than the represented subject, it is tacitly assumed that a useful description method should yield compact representations in order to facilitate the recognition and understanding of higher level organization. The noise component of the input has no structure or its structure is not recognizable at the given processing level in the hierarchical system. In contrast, information has structure and in turn, it can be compressed by highlighting the structure.

What if a novel input arrives? Novel input, by definition, has two properties: (i) the input may have structure, (ii) the input and the embedded structure has not yet been encountered by the reconstruction architecture. The bottom-up transformation channels let through the experienced structural components and filter the non-experienced structural component of the input. The filtering results in slow reconstruction, which is in contrast to the case of familiar

(learned) inputs, when reconstruction is fast. After adjusting the bottom-up transformation to enable the transmission of the structure in the novel input, reconstruction becomes faster.

Reconstruction is perfect, if the bottom-up transformed input creates an internal representation, which is able to reconstruct the structural part of the input. This latter process requires the tuning of the top-down transformation, too. In the case of perfect reconstruction, bottom-up and top-down transformations invert each other and no error correction and no iteration are needed. The system's functioning can be seen in a feed-forward manner, where top-down transformation simply reinforces the bottom-up one. We thus conclude that – in reconstruction networks – familiarity and novelty are tied to reconstruction speed.

## Neurobiological consequences

The model was successfully mapped onto the hippocampus and the adjacent medial temporal lobe structures (Lőrincz and Buzsáki, 2000, Lőrincz et al., 2002a, b). This region is thought to be responsible for higher order memory (re)-organization. One implicit evidence is that lesion to this area may give rise to anterograde amnesia in which the ability of learning new things is impaired, whereas past memories are typically spared (Knowlton and Squire, 1994). The novelty of our mapping was that starting from a relatively small set of hypotheses many structural and functional features could be derived. These results may be considered as indirect predictions of the model. Without further assumptions, we could also show the emergence of some specific low order memory functions. One intriguing example is a functional explanation of a specific category learning disorder exhibited by patients of Alzheimer's disease (Kéri et al., 2002). In our model, we could also demonstrate by means of simulations the inherent connection between repetition suppression observed in neural activity upon repeated presentation of external stimuli *and* priming (Szirtes and Lőrincz, 2002, Lőrincz et al., 2002b), which is a long-suspected relationship (Miller and Desimone, 1994). A few direct and falsifying predictions could also be made. These predictions have been reinforced recently: In accord with the model's prediction on temporal properties, one specific feature (large and tunable temporal delaying capabilities) of a peculiar sub-region (the dentate gyrus) has been observed since then (Henze et al., 2002). Another prediction of the model concerns temporal integration (that is accumulation of information during a longer time interval) at the internal representation level. Temporal integration is exhibited, e.g., by the maintenance of spiking after the excitation stops. This property has been found recently in the deep layers of the entorhinal cortex (Egorov, et al., 2002)., which is exactly the container of internal representation suggested by our model. This property can be contrasted to other layers (e.g., the superficial layer of the entorhinal cortex) where activity self-terminates, as it was demonstrated by the same work.

All these are preliminary results requiring further investigations. We are also aware that resolving the fallacy will not explain consciousness. On top of that, it is possible that this single issue may not even be on the right track in understanding the nature of consciousness. And yet, the method to follow structural considerations while dropping anthropomorphic features has already lead to testable predictions.

Summarizing our view, instead of being trapped by the Turing test we suggest seeking other methods for investigating consciousness. We think that armed with the Turing test alone, makes us too heavy to move on for two fundamental reasons:

1. The Turing test is neither necessary nor sufficient to establish if "there is a ghost in the machine".
2. The Turing test is not capable to provide testable predictions about beings with or without consciousness. It is challenging to think of better alternatives.

Acknowledgements
We are most grateful to Stevan Harnad for his helpful comments and, in particular, pointing out the importance of the mirroring problem. This work was supported by the Hungarian National Science Foundation under Grant No. OTKA 32487.